\documentclass[journal]{IEEEtran}
\usepackage{cite}
\usepackage{stackrel}
\usepackage{graphicx}
\usepackage{subfigure}
\usepackage{caption}
\usepackage{amsmath}
\usepackage{amsfonts}
\usepackage{xfrac}
\usepackage{algorithmic}
\usepackage{array}
\usepackage{stfloats}
\usepackage{multirow}
\usepackage{color}
\usepackage{tabulary}
\usepackage{booktabs}
\usepackage[usenames,dvipsnames]{xcolor}
\usepackage{colortbl}
\usepackage{eurosym}
\usepackage{pbox}
\usepackage{threeparttable}

\begin{document}

\title{Business Case and Technology Analysis for 5G Low Latency Applications}
\author{\IEEEauthorblockN{Maria A. Lema,
Andres Laya,
Toktam Mahmoodi,\\
Maria Cuevas,
Joachim Sachs,
Jan Markendahl and
Mischa Dohler}

\thanks{Maria A. Lema, Toktam Mahmoodi and Mischa Dohler are with the Cegmntre for Telecommunications Research, Dept. of Informatics of King's College London, U.K. (email:\{maria.lema\_rosas, toktam.mahmoodi, mischa.dohler\}@kcl.ac.uk).

Joachim Sachs is with Ericsson Research, Sweden (email: joachim.sachs@ericsson.com)

Andres Laya and Jan Markendahl are with KTH Royal Institute of Technology, Sweden. (email:\{laya, janmar\}@kth.se)

Maria Cuevas is with British Telecom, U.K. (email: maria.a.cuevas@bt.com)}}

\maketitle

\begin{abstract}
A large number of new consumer and industrial applications are likely to change the classic operator's business models and provide a wide range of new markets to enter. This article analyses the most relevant 5G use cases that require ultra-low latency, from both technical and business perspectives. Low latency services pose challenging requirements to the network, and to fulfill them operators need to invest in costly changes in their network. In this sense, it is not clear whether such investments are going to be amortized with these new business models. In light of this, specific applications and requirements are described and the potential market benefits for operators are analysed. Conclusions show that operators have clear opportunities to add value and position themselves strongly with the increasing number of services to be provided by 5G.
\end{abstract}

\begin{IEEEkeywords}
market drivers, use cases, business models, low latency, Tactile Internet
\end{IEEEkeywords}

\maketitle

\section{Introduction}

During the past decades, introducing evolutions of mobile communications systems in the consumer market was mainly driven by improving the quality of mobile broadband services. These services are generally bounded by the use of smart phones and tablets which generate high amounts of data in the form of voice and video. Having nearly every device connected to the network significantly boosts the number of applications that run through the Internet. In fact, the future Internet is already meant to provide a higher number of services targeted to satisfy both consumer and industry needs, the so-called verticals within the fifth generation (5G) mobile network. The integration of healthcare, industrial processes, transport services or entertainment applications on one hand generates new business opportunities for the network operators, and on the other hand poses strong requirements to current network deployments, which opens the door to the design of the future 5G network. Making one single communications network, capable of delivering all services across multiple industries is in a sense very challenging, since every use case or application will need to fulfill different requirements.

Probably the most disruptive change in nowadays internet is the support for remote real-time fully immersive applications. The Tactile Internet will extend current transmission capabilities of voice and video to touch and skills. The transmission of multi-sensorial signals (including the sense of touch, i.e., haptics) can well improve the immersiveness and the overall experience. Real-time remote interaction is often based on action and reaction, as for example force feedback haptic signals or virtual and augmented reality. These closed loop transmissions limit the round trip latencies, since large delays between the action and reaction may cause instability of the control loops and impair synchronism of different data flows. Applications that require ultra-low latency networks across the different industries are summarised in Table \ref{tab_5G_capabilities}. Note that \emph{Smart Energy} is also an important use case when evaluating low latency applications, however it is left out of this study due to the lack of primary data sources.

\begin{table}
	\centering
	\caption{Low Latency Applications in 5G Networks}
	\label{tab_5G_capabilities}
    \begin{threeparttable}
	\begin{tabular}{l l}
		\toprule
		\textbf{Industry Vertical} 	& \textbf{Application} 	\\
		\midrule
		\midrule
		\multirow{3}{*}{Healthcare Industry} &
		Remote robotic surgery with haptic feedback \\
		& Remote diagnosis with haptic feedback \\
		& Emergency response in ambulance	 \\
		\midrule
		\multirow{4}{*}{Transport Industry}\tnote{1} &
		Driver assistance applications\\
		& Enhanced Safety \\
		& Self-driving cars\\
        & Traffic Management\\
		\midrule
		\multirow{2}{*}{Entertainment Industry} &
		Immersive entertainment\\
		& Online gaming \\
		\midrule
		\multirow{2}{*}{Manufacturing Industry} &
		Motion Control \\
		& Remote Control with AR Applications \\
		\bottomrule
	\end{tabular}
                \begin{tablenotes}
                    \item[1] This industry includes automotive, public transport and infrastructure.
                \end{tablenotes}
    \end{threeparttable}
\end{table}

The telecom community has been working on defining several technological solutions that allows to broaden the network services scope and include all these new applications. In this sense, there is a clear technological roadmap for 5G where the support for low latency is in the agenda. Ultra-reliable low latency service delivery is probably one of the most challenging 5G goals, which may imply costly investments. Yet, one of the open questions and main concerns in the telecom community is to answer if there is a real market need for a 5G network to support this stringent latency values. Investigations done so far in 5G shows that there is a clear interest from the industry verticals to integrate communications and technologies \cite{METIS}, however there is still not a clear picture of the potential revenues or business models for the different telecom players. We need to understand the trends of each industry vertical, and also, how likely are these new use cases to be transformed in new business models. More importantly, the telecom community needs to understand which role to play in the new service development processes of the different industry verticals. The introduction of low latency applications represents substantial technical challenges but we can also foresee major changes in the way businesses are made. Here we can learn from ongoing business research related to Internet of Things (IoT) services and 5G systems.

This article describes a number of low latency use cases being considered so far by the community and surveys the main system requirements to be fulfilled. We then discus the need for a technology evolution in order to support ultra-low latency services by analysing the current available technologies. We thereupon look at the market perspectives of each industry vertical separately, analysing the market size, the relevant stakeholders and the potential business opportunities for network operators.

We present a detailed overview of the trends, potentials and challenges related to ultra-low latency applications. Based on an extensive review of the available literature, trend reports and in-depth discussions with key stakeholders, we also elaborate on the opportunities for telecom actors. Further on, we draw parallels with the ongoing business transformation for the telecom industry that has started with the introduction of IoT solutions.

The remainder of this article is organised as follow: Section II describes in detail the methodology of this work, and explains the rationale behind the structure of this study. Moreover, Section III goes through the trends and challenges of each of the novel use cases being considered by the different industry verticals; Section IV exposes the actual performance of commercial fixed and mobile networks measured by the United Kingdom (UK) regulator \textit{Ofcom}. Section V discusses the need to introduce a new network able to satisfy the latency requirements of these new consumer and industrial applications. Section VI describes the main market trends on each industry vertical, highlighting the particular business interest in the support for low latency applications. In Section VII we discuss the business transformation, by drawing parallel lines with the IoT services. Finally, this paper is concluded in Section VIII.

\section{Methodology}
We depart from the following research question: \textit{will it be cost effective for telecom players to build ultra-low latency in 5G networks?} In order to provide an answer to this question, first we overview the industry verticals that could benefit from low-latency technologies. We review the specific applications in terms of key performance indicators and technological requirements. Then we contrast these requirements with the available communication technologies to highlight the current technological limitations, and motivate the need for a change in 5G. We finally present the implications of the cases and technologies in terms of market opportunities.

Thus, the aim is to present the match between the main trends in the technological arena and future business opportunities that will arise as a result of enabling low latency applications. The research strategy is exploratory in nature, using cumulative case research studies of promising use areas of 5G technologies and match them with technical requirements and business opportunities for telecom players.

Cumulative case studies allow the aggregation of multiple sources of data, including qualitative and quantitative sources, which provide triangulation to support the validity of findings from empirical data \cite{Davey1991,Mann2006,Eisenhardt1989}.
Based on the accumulated data, we show the current performance targets in the different industries as well as the different market trends in the context of the UK. The UK context represents a relevant area with industrial and governmental interest in low latency communications as means to advance the industrial development. In addition, the interpretation of the data sources is based on contextual understanding, providing rich and deep observations by having the core research team close to the research setting \cite{Guptaed.2015}.

The technical survey is divided into three components: performance of currently deployed networks, description of latency aspects in 4G and fixed networks and finally, survey of current trends towards low latency delivery in 5G.

\subsection{Data Collection}

We use a combination of primary and secondary sources, which include:
\begin{itemize}
\item Interviews with players of the relevant industries regarding the trends and targets in the inclusion of communication technologies. Profiles included are: robotic surgeons, market researchers and technology researchers.
\item Discussions (workshops and round tables) with main telecom players to understand how technology can deliver the verticals' main targets.
\item Collaborative research projects with experts across different industries that contribute to a co-creation process of the required technology that is capable to match the requirements of the new use cases.
\item Extensive literature review including standard contributions, regulators published data, research contributions, articles and market research studies which provide additional supportive information.
\end{itemize}

\subsection{Data analysis}
\begin{figure}[t]
  \centering
  \includegraphics[scale=0.6]{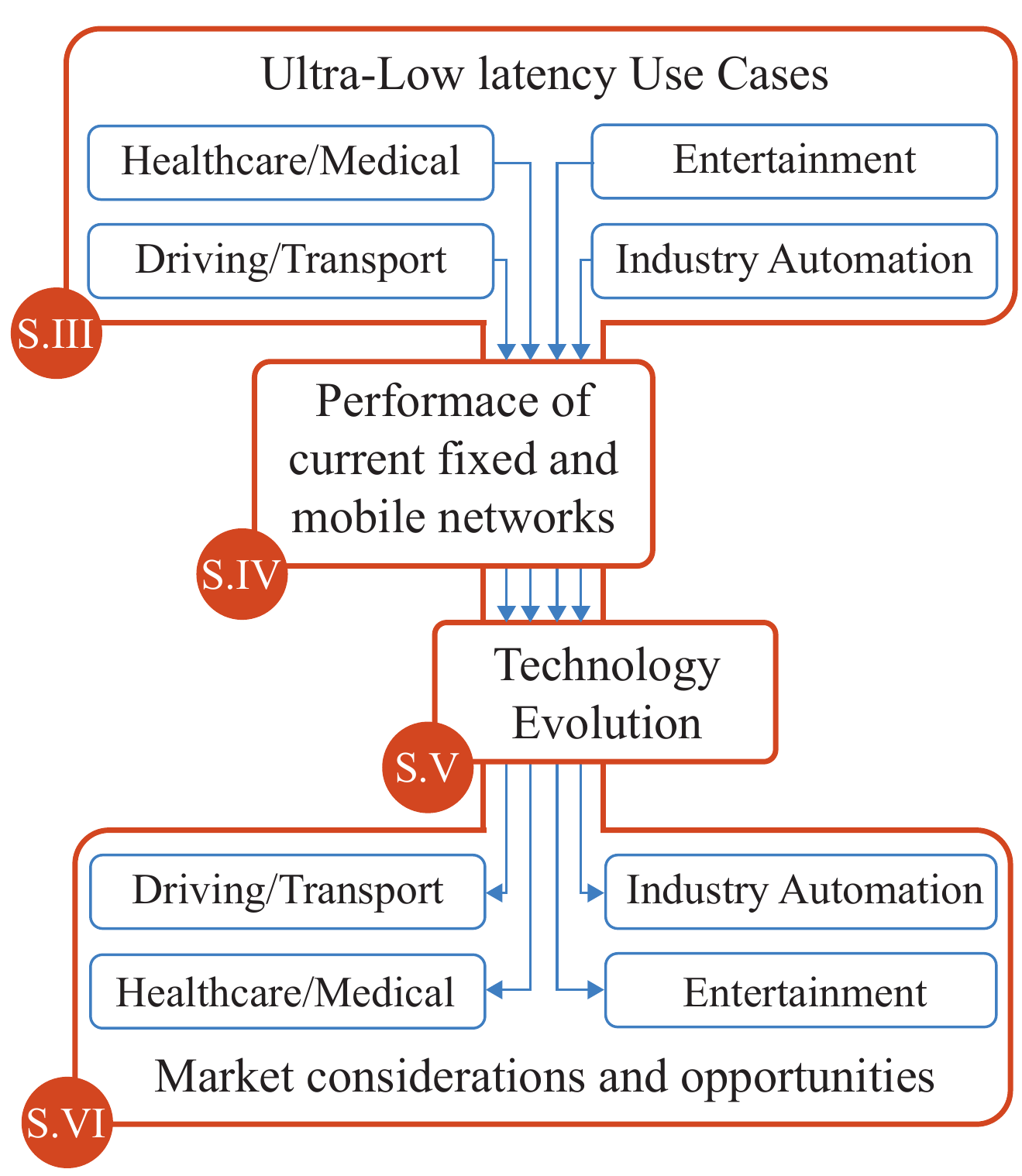}\\
  \caption{Research approach.}\label{fig:strategy}
\end{figure}

The data analysis follows the approach presented in Fig.~\ref{fig:strategy}.
First we describe the context for four uses cases where the use of ultra-low latency technologies offer great opportunities in the form of new applications or services. Then we present a quantitative analysis on the limitation of current fixed and mobile networks and further explore the technology evolution that is being realised to enable these new applications. We then make a qualitative analysis of the possible business opportunities for telecom players in the four identified cases. Finally, we present a discussion on lessons learned from the IoT business and ecosystem transformation for the telecom industry and how the convergence of different industries in changing the position of traditional telecom actors.

The objective is to relate input on novel use cases being considered by the different industry verticals with input on the actual performance of commercial fixed and mobile networks. This will evidence the immediate technological limitations that should be overcome in order to exploit business opportunities supported by low latency applications. Such opportunities are of course considered assuming the existence of the corresponding network build-out that enables them.

\section{Ultra-Low latency Use Cases}

\subsection{Remote Healthcare and Medical Intervention}
Communications networks in general, and mobile networks in particular, are the key enablers of the main targets and trends of the healthcare industry providing: cloud-based solutions that improve the accessibility of high-resolution medical data, increased capacity for real-time high-definition video transmission, support for massive number of connected devices such as e-health wearables, robust mobility support, and finally ultra-low latency communications. To this end, two main applications considered in the 5G context are: ($i$) remote healthcare and precision medicine with the use of bio-connectivity, and ($ii$) remote intervention with the use of remote robotic surgery \cite{5GPPWPMedical}.

In the bio-connectivity context, there is a trend for decentralization of hospitals, where medical care can be provided at home or on the move (i.e., emergency response in ambulances), electronic medical records and data analysis for predictive healthcare, as well as the use of embedded systems to perform individual pharmaceutical analysis. In the remote surgery context, the aim is to break the obstacle of geographical boundaries in providing high quality healthcare in the most complex medical interventions and surgeries.

Revolution in the healthcare industry is already underway by integration of connectivity in the sector and can be seen in the following avenues \cite{AccentureWP}:
\begin{itemize}
\item Integrated systems that combine medical records with different communication methods, remote care and process management.
\item Redirection of interventions from expensive hospitals with the use of tele-medicine, remote care and mobile care.
\item Engagement of society with the use of wearables that bring advantages such as collaborative and shared decision making, chronic patient monitoring and management.
\item Personalized treatment and plans that enable to coordinate care, target resources and improve health outcomes.
\end{itemize}

The above use cases or applications can be clearly separated into those that are latency dependent and those that are not. Medical interventions require lower latency than remote care, where evolution over time is an important aspect rather the instantaneous measurement or events.

In remote interventions, the level of interaction of the medical expert will determine the latency tolerance of the system. In a \textit{tele-mentoring} scenario, which refers to guidance of one health-care professional by another, the level of interaction can vary from verbal guidance while watching a real-time video stream, to taking control over the assistant via a robotic arm. On the other hand, in a remote surgery scenario, or \textit{tele-surgery}, the entire procedure is controlled by a surgeon at a remote site. Current surgical equipment is not equipped with tactile sensors that can allow the doctor to feel the stiffness, therefore the doctor has completely lost the sense of touch by replacing his hands by a robotic arm.
In past tele-surgery trials, with no use of haptic feedback, have shown that doctors can compensate high levels of latency, depending on their level of interaction. In the tele-mentoring context, mentors can compensate delays up to 700\,ms in less interactive scenarios, while in more interactive mentoring scenarios a shorter delay of up to 250\,ms is required. In the context of tele-surgery, real experiments have determined that the maximum tolerable delay is 150\,ms \cite{BJUBJU4475}. Since there's no haptic feedback included, the reported delays here are all one way delay.

However, depriving the surgeon from the sense of touch, impedes the doctor to fully exploit its palpation skills. Haptic feedback in remote surgery and diagnosis scenarios can increase the accuracy in detection of cancer nodules, for example. Hence, the robotic community together with the medical one are working towards the inclusion of such sensors that can allow an efficient use of the medical palpation \cite{6819407}. Adding the haptic feedback, however, tighten the requirements on latency, since kineasthetic devices work in closed control loops and the two ends (action and reaction) should operate in sync with each other. Previously reported figures show that, tele-surgery in the presence of haptic feedback requires end to end round trip times (RTTs) of lower than 10\,ms \cite{10.1109/9.24201, 10.1177/027836499201100204}.

A number of business initiatives have started in this area, including:
\begin{itemize}
  \item Verily, a subsidiary of Alphabet Inc. just announced the creation of its surgical robot division: Verb Surgical will integrate technologies such as advanced imaging, data analysis, and machine learning to enable greater efficiency and improved outcomes across a wide range of surgical procedures \cite{IEEESpectrum}.
  \item SRI’s Research prototype tele-robotic surgical system, M7: Auditory, visual, and tactile sensations, including the force or pressure felt while making an incision, are communicated directly to the surgeon performing the operation, and also includes motion compensation for operating in a moving vehicle.
 \item RAVEN and RAVEN-II are surgical robot platforms for research, to improve performance and capabilities of tele-operated surgical robots. The main objective is to provide a common open platform software and hardware to support research innovations \cite{6363574}.
  \item National Taipei University of Technology (NTUT) and Taichung Hospital are working on an ambulance-support Emergency Response system. A wireless sensor system transmits patient parameters to the emergency wards of hospitals while patients are in the ambulance on the way to the hospital \cite{Chang2010}.
\end{itemize}
Other state of the art developments in the context of robotic surgery can also be found in \cite{RoboticSurgery}.

\subsection{Assisted Driving and Transport Services}
Following the definition in \cite{GSMAAutomotive} the term intelligent transport systems (ITS) refers to the use of Information Technology, sensors and communications in transport applications, aiming at providing more efficient movement and seamless journeys for people in both public and private means of transportation. In particular, the automotive market is transitioning to a fully connected car, which empowers new user experiences such as autonomous or assisted driving to increase safety, reduce pollution and congestion \footnote{Autonomous driving does not require human interaction and Assisted driving is to enhance the Human driving experience with additional information.}. Also, the use of traditional internet services, such as video and music streaming, which will allow extending the smart phone applications inside the transport services, encompassing high definition video streaming, real-time video streaming or low latency applications as cloud gaming \cite{3gpp.22.891}.

Several applications and use cases are already under research and development, the most representative being \cite{5GPPPAutomotiveWP}:
\begin{itemize}
\item Automated driving, which increases the level of automation when driving. Cars benefit from local information transmitted from other vehicles or the infrastructure, this allows to better adapt to the traffic situation. Some popular applications in this context are automated overtake, cooperative collision avoidance and high density platooning.
\item Road safety and traffic efficiency services are introduced mainly to increase driver’s awareness. This use case relies on the principle that the connected car is constantly sending or receiving information (status or event) to the infrastructure, other vehicles or the network. Automakers are considering applications such as see-through other vehicles, vulnerable road user discovery, bird’s eye view (in intersections mainly).
\item Digitization and transport logistics which aim to collect traffic information and use it cooperatively to improve route optimization, energy consumption and travel times transport systems.
\item Intelligent navigation systems are introduced to enhance the experience with augmented reality and real-time video of traffic information.
\item Information and entertainment (i.e., infotainment) services being on the move.
\item Nomadic nodes is an application to improve capacity and coverage and considers the vehicle as a small cell while parked.
\end{itemize}
It is certain that mobile communications are key enablers of a majority of the potential future automotive user applications. One of the major challenges is the massive number of devices that are going to be constantly accessing the network, mainly in machine type communications (MTC), since both vehicles and infrastructures are going to be filled with sensors and actuators. Most of these devices require very low delay and ultra-high reliability, known as critical MTC communications. However, given the diversity of applications addressed in the automotive sector, different level of latency support for such wide range of applications should be considered. For instance automated overtaking systems require a maximum tolerable end to end latency of approximately 10\,ms on each message exchange. When video is integrated as in the see-through application described in \cite{VW}, encoding and decoding video would suppose prohibitive delays, thus very high data rates are required to transmit real-time raw video. For a 30\,frames per second video feed a capacity of 220\,Mbps and an end to end latency of 50\,ms shall be supported. In such scenarios it is important that the communications system undertakes an overhead reduction in MTC, since a high number of control signals are often generated with respect to the amount of data.

Also, given that mobility is the protagonist, enhanced solutions in terms of handovers and device discovery for high capacity and low latency communications are required, as well as fast recovery processes after coverage loss. To do so, it is required to allow easy integration of multiple radio technologies to provide seamless experience and efficient use of resources between different technologies. Other technological solutions that need to be considered are those related to active quality of service (QoS) management, where the network can constantly monitor the level of experience and trigger changes in the network when KPIs cannot be satisfied. An example of this is the inclusion of network controlled device to device (D2D) communication, where the network can shorten the end to end path in an MTC context.

The automotive industry and other players have already seen the business opportunities, and current commercial solutions go from smart phone applications that share real time information among commuters and drivers, to integrate the IoT in the infrastructure and adding more intelligence to the car with the use of communication systems. Some examples are:
\begin{itemize}
\item Smart commuting applications that offer personalized commuting services,
\item \emph{Tranquilien}: applies big data and analytics to successfully predict how crowded a train will be throughout the day and up to a week in advance, changes people behaviour and works with 85\% accuracy.
\item \textit{FordPass} mobile App offers services that range from finding and paying for parking, borrowing and sharing vehicles, location of services or enhanced mobility.
\item Auto manufacturers such as Ford, Audi, Jaguar Land Rover or Volkswagen are working towards the assisted driving use case.
\item Google and Apple have their own business line for self-driving cars. Google self-driving car accident has engaged the discussion of the need for Vehicle-to-Vehicle (V2V) communications to support automated driving.
\item Uber has strong agreements with Universities and research centres for self-driving car research.
\end{itemize}

\subsection{Entertainment: Content Delivery and Gaming}
The media and entertainment business is already undergoing big changes which are mainly caused by the behavioural change of individuals. We are no longer mere consumers and we actively interact during the media and entertainment enjoyment \cite{5GPPWPMedia}.

Changes in the entertainment industry already started with the increase of availability of high speed networks, both mobile and fixed internet. These together with data centres and cloud computing, have contributed in large extent to the increase of more immersive experience demand. Consumers in general have been largely influenced by the growing capabilities of the devices together with the innovative services provided by the different content generators, and people prefer to watch streamed on demand video and television rather than scheduled programs \cite{TVandMedia}. Broadly speaking, video on the go, streaming services at home, live events experiences and more entertainment services have become great socio-economic drivers for the entertainment business.

Specific applications being considered in this sectors can be classified as follows \cite{5GPPWPMedia}:
\begin{itemize}
\item Ultra-high fidelity media: high immersive viewing experience in both live and streamed content, and in all kind of devices and locations.
\item On-site live event experience: improve the live experience on site by offering better experience to the customer (the audience of a live event is enriched with replays, choice of cameras, or integrated augmented reality).
\item Immersive and integrated media: immersive and interactive media consumption, with smart adaptation to the ambient of viewing including 3D video transmission.
\item Cooperative media production: content captured and shared immediately, which provides immediacy access to the content.
\item Collaborative gaming: full immersive multi-sensorial experience, moving from home based experience to anywhere.
\end{itemize}

Full immersive entertaining experiences, such as gaming, have been enriched with increased graphic resolution and simultaneous events happening among different active users, which makes \emph{gamers} want as much realism as possible and also wish to have the most immersive experience while playing. And currently, the gaming industry's efforts are placed on enhancing gamer experience by adding elements of virtual or augmented reality (VR, AR) and bio-sensing, which allows the player to detect people in the game in real or imaginary worlds; it also adds the capability of motion capture to interact with objects surrounding with realistic force feedback. AR is expected to revolutionize the gaming industry in general, because of its inherent realism as it lets the gamer experience real world in tandem with the game played.

In this sense, an ultra-reliable low latency network capable of providing the fully immersive multi-sensorial services, through video, audio and tactile can further enhance the consumer experience, in both content delivery and gaming. Some of the main technical challenges to deliver these new kind of entertaining services closely resemble to other use cases already described, QoS guarantee in the form of: data-rate, mobility, end to end latency, coverage and reliability. However, the main stringent requirements to provide full immersive experience are related to the augmented and virtual reality, which limits very much the allowed end to end latency as it is fundamental to deliver good experiences and extend the realism of the game in a simulated or a real environment. For virtual reality and augmented reality 15\,ms to 7\,ms application to application delay, i.e., action to reaction, is the threshold to provide a smooth action-reaction experience. The maximum allowed latency determines the level of capacity, since encoding and compression takes a huge part of the latency budget, VR content delivery can be very demanding in terms of capacity as well. In general, the video or image related to the direction of where the user is looking is actively transmitted. To reduce the processing burden in the device end, 5G should integrate high processing in mobile edge computing clouds.

Immersive content delivery and gaming is gaining huge interest in both research and industry, mainly driven by its stringent latency and capacity requirements. Several solutions related to the previously described use cases are fully implemented, and many other are still in developing process, some examples are:
\begin{itemize}
\item Cast it on the TV applications, such as Google Chromecast
\item OculusVR and many other VR/AR interfaces such as Samsung, Google, HTC Vive
\item VR applications and companies are boosting: broadcast sports (Fox Sports with NextVR), immersive movies, 3D selfie to buy clothes online, see items in real size, for real estate and tourism.
\item New interfaces such as Gloves that add haptic capabilities.
\end{itemize}
\subsection{Industry Automation}
The factory of the future (FoF), or Industry 4.0, is a European Commission vision to re-industrialize Europe on how manufacturing process will be operated in the future. There is a strong trend from the European Union to digitise the industry to provide higher value products and processes \cite{EUWhitePaper}, and European organisations have recognised that the path towards the forth industrial revolution consists of intelligent networking of product development and production, logistics and customers. In this sense, FoF are indeed not stand alone closed entities, but will be a part of a larger value chain and ecosystem.

Until today the IoT has provided powerful solutions to improve industrial systems and applications. In the past, industrial monitoring tasks could be carried out with the use of wireless sensor networks (WSNs), which interconnected a number of intelligent sensors to perform sensing and monitoring \cite{10.1109/COMST.2015.2444095}. The evolution of WSNs has largely contributed to the development of IoT in mobile communications, and industrial applications nowadays include much more than monitoring and tracking and can integrate all these added capabilities intended for the digitalization of FoF.

Thus, the main competitive trend in the manufacturing business is to evolve into intelligently connected production information systems that can operate beyond the factory premises, and in this context several applications considered range from time-critical operations to remote control of factory equipment \cite{5GPPWPIndustry}:
\begin{itemize}
\item Time critical process optimization and control: real-time optimization based on instantly received information from monitoring or interaction between different operators, remote control of robotic operations and
collaborative robots in closed-loop control systems. This use case family is characterized by communication latencies that may go below 1\,ms.
\item Non-time critical communications, encompasses applications such as non-critical localization of assets and goods, quality control and sensor data collection.
\item Remote control applications mainly based on augmented reality to provide support in production and maintenance.
\item Seamless communications along the value chain providing connectivity between different production sites and other parties.
\end{itemize}
To effectively support real-time cooperation and intervention the network shall deliver flexible and converged connectivity that ensures a seamless experience across multiple mediums, such as wired and fixed networks, multiple vendors and multiple technologies. In this sense, the network needs to provide a highly heterogeneous multi-connectivity scenario, where everything is capable of communicating even in harsh industrial environments. Also it is necessary a fast and reliable reconfiguration of QoS and traffic demands, to enable fast network adaptation to current application's needs. As for the network support for delay requirements, apart from stringent latency, the jitter needs to be contained as well. Some applications involve high data transmission with the use of wearables, for instance 3D video or augmented reality content, other applications involve low data transmitted by the different sensors. As well, an underlying requirement is the efficient management of MTC.

The manufacturing business started adopting technological innovation long ago with the inclusion of WSN, therefore there are commercial options in the market as examples of the evolution towards the industry 4.0  \cite{EUWhitePaper}:
\begin{itemize}
\item \textit{Trelleborg}: co-bots (collaborative robots) that can work next to people thanks to improved sensors, which shut the robot down if someone gets too close.
\item Adidas has developed mass customization for sports shoes involving digital design and 3D printing.
\item Siemens has set up a showcase electronics factory based on fully automated and networked production.
\item Strong companies such as Airbus have research roadmaps towards the FoF, considering plug and play robots, enhanced simulation tools with the use of VR and integrated production.
\item Fujitsu has already provided some solutions that improve the on-site working conditions with the use of AR technology \cite{FujitsuWP}.
\end{itemize}
\section{Performance of Existing Network Solutions}
Previous sections have focused on the latency requirements for the new use cases and applications to be integrated as services in the next generation of mobile communications. Some applications require close to real time response from the network (i.e., less than 5-10\,ms) and pose figures that are a real challenge for today's network deployments. In order to move on into 5G, and assess the real need for changes in the actual network, it is necessary to evaluate how today's networks perform in terms of latency. Since some of the use cases presented do not necessarily require mobility all the time and can be carried out in fixed-line broadband network environments, such networks have also been considered in the analysis.

\subsection{4G and 3G Network Performance}
\begin{table*}
	\centering
	\caption{RTT in 3G and 4G Mobile Networks (original data from \emph{Ofcom})}
	\label{tab:Mobile_Performance}
	\begin{tabular}{l l l}
		\toprule
		\textbf{Performance indicator} 	& \textbf{3G} & \textbf{4G} 	\\
		\midrule
        \midrule
		Average RTT & 63.5\,ms & 53.1\,ms \\
Lowest RTT & 58.9\,ms & 49.8\,ms \\
Distribution of RTT & 43.5\% between 20 and 60\,ms & 68.9\% between 20 and 60\,ms	\\
		\bottomrule
	\end{tabular}
\end{table*}

The numbers discussed in the following lines are based on a research study on mobile broadband performance done by the UK regulator \emph{Ofcom}, and all performance results are collected in \cite{OfcomMobile}. Since the focus of this work is latency, we just consider the delay figures presented in the report. The latency is defined as the responsiveness of the network, and it is measured as the delay of transferring data to and from the user equipment (UE). For this particular tests, the latency of the mobile network was measured doing a ping test, the resulting time span corresponds to the round trip time (RTT). The reader is referred to \cite{OfcomMobile} for further details on the performance tests; the main findings related to the latency values are summarised in Table \ref{tab:Mobile_Performance}. Results show that 4G is much more stable with latency than 3G systems, however, much work needs to be done to support the low latency figures discussed previously for new real-time services in mobile networks.

\subsection{Fixed Broadband Network Performance}

Similarly, in the case of fixed networks the performance metrics are based on an Ofcom report on the performance of UK fixed networks presented in \cite{OfcomFixed}. As well, this study is very broad in terms of network performance and several figures are provided. For the sake of the interest of our study we only focus on the latency measurements. Latency is measured following a similar approach as in the mobile network tests, defined as the time for a single packet to travel from the user computer to a third-party server and back, i.e., RTT. Also related with delay performance, we summarise the jitter performance figures which is defined as the rate of change of latency.

\begin{figure*}
  \centering
  \captionsetup{justification=centering}
  \includegraphics{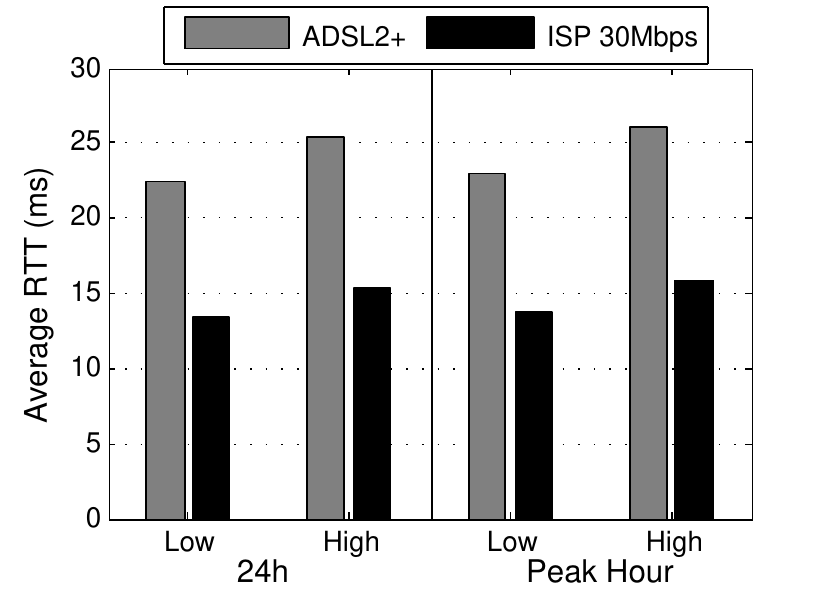}\\
  \caption{RTT performance of Fixed Networks (original data from \cite{OfcomFixed}). \emph{Low} and \emph{high} correspond to the average lower and higher experimental values}\label{fig:latency}
\end{figure*}

\begin{figure*}
  \centering
  \captionsetup{justification=centering}
  \includegraphics{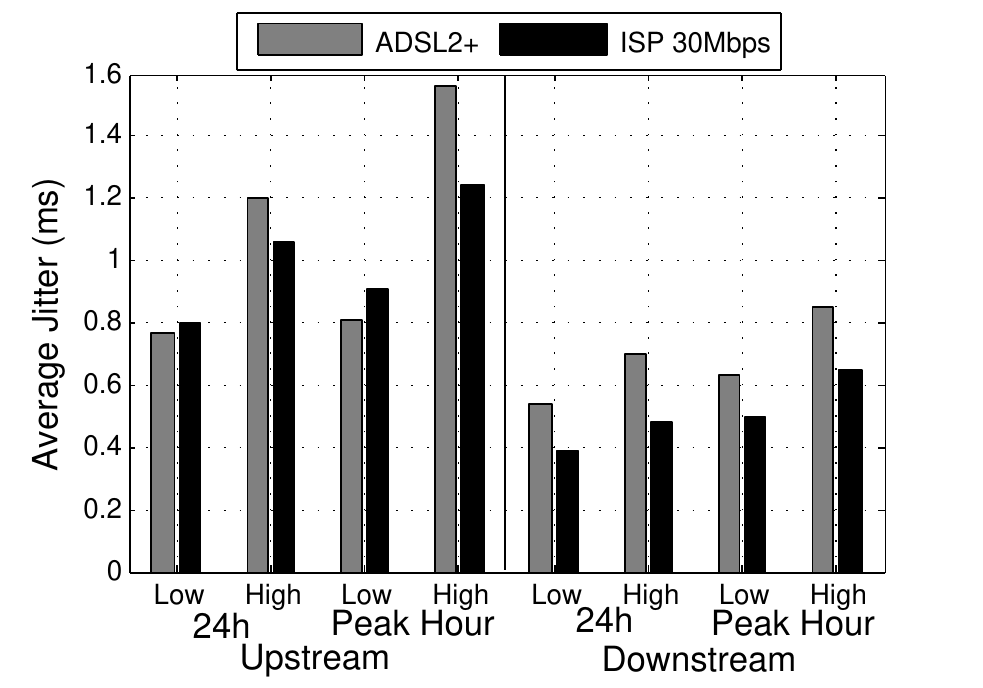}\\
  \caption{Jitter performance of Fixed Networks (original data from \cite{OfcomFixed}). \emph{Low} and \emph{high} correspond to the average lower and higher experimental values}\label{fig:jitter}
\end{figure*}

Key findings for fixed network RTT values are summarised in Fig.~\ref{fig:latency} and \ref{fig:jitter}. We show the results for ADSL 2+ and Fibre services of more than 30\,Mbps. Fixed networks present significant performance differences in the jitter during different times of the day, and it is very sensitive to congestion. This is an important limitation for fixed networks, that can impact the support for low latency services that do not necessarily require mobility.

\section{Technology Evolution}

Safety related applications in the automotive industry require end to end delays as little as 10\,ms, and robotic tele-operations (surgery or remote control of robots in any other application) that require haptic information feedback need to have stable latencies that are below the round trip values of 10\,ms. Augmented and virtual reality use cases require even lower latencies, round trip latencies as low as 7\,ms to have a full immersive experience. Based on the latency requirements analysis, and the current performance of 3G, 4G and fixed networks, it is clear that some changes need to be made to the current network solutions in order to successfully incorporate these new applications.

\subsection{Overview of Data Transmission Delay Components}

\textbf{Current 4G networks:}

Delay components of the 4G standard Long Term Evolution (LTE) in both uplink and downlink are carefully analysed and quantified in \cite{3gpp.36.881}. In particular, the main delay components for a data transmission from the user equipment (UE) to the packet gateway (PGW) (Uplink) and back (Downlink) are:
\begin{itemize}
\item \textit{Grant Acquisition Process} Once the UE has created and packetized data this is ready to transmit towards the base station, or evolved Node B (eNB). To do so, the UE must send a \emph{Scheduling Request (SR)} message to the eNB, which the eNB will answer allocating the user a scheduling grant (i.e., its transmission opportunity or slot) to transmit the packetized data. It is important to note that the UE must send the SR in a valid control channel: Physical Uplink Control Channel (PUCCH). Once the UE receives and decodes the scheduling grant the packetized transmission takes place in the Physical Uplink Shared Channel (PUSCH). One of the main limitations of this resource allocation method is that the UE needs to wait until a valid PUCCH is available, and this depends on its periodicity. Low periodicity increases delay in waiting for SR, and high periodicity increases the control overhead. Assuming the LTE Release~8 functionality the average waiting time for a PUCCH at a periodicity of 10\,ms is 5\,ms.
\item \textit{Random Access Procedure} The Grant Acquisition process is useful for UEs that are already connected and aligned with the eNB. When this is not the case, or there is no control channel PUCCH configured for the SR, the UE needs to initiate a random access channel (RACH) process, which also serves as an UL grant acquisition method, and it is known as random access based scheduling request. The total process of RACH procedure including RACH scheduling period, preamble detection and transmission and both processing (UE and eNB) delays is 9.5\,ms. Note that when the user is not synchronized and performs random access scheduling request it is no longer required to send a SR in the PUCCH.
\item \textit{Transmission Time Interval (TTI)} Every transmission (request, grant or data) is done in a subframe with 1\,ms duration. Hence, it is the minimum transmission unit in LTE.
\item \textit{Processing} It is proportional to the transport block size for data processing. In general this value is considered to be 3\,ms for PUCCH and 5\,ms in the UE side for the RACH processing and UL alignment.
\item \textit{Packet Retransmissions} In the uplink hybrid automatic repeat request (HARQ) process round trip time is 8\,ms (for frequency division duplex) due to its synchronous nature; in the DL it is not directly specified as the scheme is asynchronous.
\item \textit{Core Network/Internet} packets queued due to congestion.
\end{itemize}

According to these latency components, the overall radio access delay quantification for an uplink transmission, can be as high as 17\,ms when the UE is  time aligned with the eNB (i.e., using SR and not RACH process) and not considering any retransmission. In the downlink transmission, since the user does not need to undergo a grant acquisition process, the overall radio access delay is 7.5\,ms, and it is mainly due to processing and transmission delays.

\textbf{Fixed networks:}

Similarly, each step of the end-to-end delay within a fixed network transmission:
\begin{itemize}
\item \textit{Propagation delay}, which is distance and medium dependent.
\item \textit{Serialization delay}, in current high speed broadband networks this delay can be in the order of microseconds.
\item \textit{Protocol delay}, connection oriented protocols such as TCP can increase the end-to-end delay with retransmission and congestion avoidance mechanisms, for real-time communications UDP is commonly used.
\item \textit{Switches and Routers}, high performance routers and switches can add 200\,microseconds, it generally contributes with around 5\% to the end-to-end delay.
\item \textit{Queuing and Buffer Management}, which is load dependent, it can add up to 20\,ms of latency.
\end{itemize}
\subsection{Latency reduction techniques for mobile and fixed networks}
By analysing the latency budget of current LTE transmissions, it is clear that some of the bottle neck points are in the low efficiency of the protocols itself, having high delay grant acquisition process and retransmissions, especially in the uplink. Therefore, solutions in reducing the complexity in this line are currently being investigated by the Third Generation Partnership Project (3GPP). Work in \cite{Lat1} shows that it is feasible to obtain very low error rates and low latencies by using a simple air interface design. Authors also underline that it is important to provide enough transmission opportunities (i.e., scheduling grants) and shorten the transmission intervals (i.e., the TTI).

The 3GPP has studied options for LTE latency reduction in \cite{3gpp.36.881}, which are being pursued in the ongoing standardization. A work item \cite{A} has just been finished, in which a fast uplink access has been defined.  The concept of semi persistent scheduling (SPS) has been enhanced to allow pre-scheduling of uplink resources with a periodicity of 1ms. Another work item \cite{B} is ongoing, which specifies a short TTI that has a length of either 2 or 7\,OFDM\footnote{Orthogonal Frequency Division Multiplexing} symbols, which corresponds to ~140\,$\mu$s or 500\,$\mu$s respectively. In addition, reduced processing times and asynchronous uplink HARQ operation are being specified.

With these features of LTE releases~14 and 15, the LTE  radio access network (RAN) latency can reach down to the millisecond level with appropriate configuration.
As a next step, a contention based uplink has been proposed (to increase the spectral efficiency), as well as features for increased reliability \cite{C}.
For the 5G new radio (NR) interface, a requirement has been defined, to be able to reach down to 0.5ms one-way 5G-RAN latency \cite{3gpp.38.913}. Within the NR study item \cite{NR}, the following concepts contribute to this goal: the flexible OFDM numerology with different sub-carrier spacing allows configurations with short OFDM symbol duration and transmission time intervals. This is complemented with a shortened slot structure and reduced processing times. In addition, several features to improve the reliability are defined.

Besides the efforts of reducing the complexity in the LTE radio access protocols, it is necessary to manage both control and data planes end-to-end paths. Some use cases, particularly those that involve communications among two close devices could well benefit from direct device transmissions, benefiting from techniques such as D2D communications. It is particularly highlighted by the 3GPP as an operational requirement of 5G to support direct communication among devices within the same network \cite{3gpp.22.891}. The main benefit of this direct communication is that transmission can be terminated much closer, without having to send information through the Core Network or the Internet. Since direct communication can be challenging in terms of frequency resource management and interference coordination, network support is essential, and eNBs could intelligently manage these type of short-range transmissions. Also, 5G should introduce some active QoS monitoring and management techniques, as part of an end-to-end QoS framework. Software defined network (SDN) type of procedures, such as the use of path computation engines that allow to measure the end-to-end latency of a particular network path that combined with the use of flow preemption techniques can certainly improve the overall latency and jitter experience.

Very much related to the end-to-end path is the placement of the network functions or content servers. For example, a user with a high mobility pattern needs mobility management entities to be close to the edge, so handover functions can be done faster and reducing load in the backhaul network. Another example is the position of data and control anchoring points. If one network path is experiencing higher latency, or a network element is increasing its load, it would be beneficial to allow the network to change the anchoring point to another device in the network that can satisfy the application's requirement better. Not all services require the same network functions, so 5G needs to support network modularisation and granularity so that a dedicated set of functions can be selected. Network function selection is one of the core the concepts of network slicing and virtualisation, the 3GPP has already considered the use of dedicated core networks, and network function virtualization (NFV) in the core of LTE has been widely studied in the literature. However, the specific set of functionalities for low latency use cases and their optimized location in the network are still undefined.

Finally, 5G needs to be a fully converged network where different access types can be intelligently selected. Multi-connectivity has been a well-researched topic in later 4G releases (Release 13 and 14 \cite{13}) and has shown good performance results when flexible network associations are allowed \cite{7582426}. Hybrid access with the use of fixed and mobile networks can also enhance latency over the network, by reducing the impact of the LTE-A access delay.
\section{Market Considerations for Ultra-Low Latency Applications}
Mobile operators, vendors and new entrants in the telecommunication industry will work on the development of the next generation of mobile networks in order to support new services with ultra-low latency requirements. The interest generated by different industry verticals should be matched with the value that telecommunications companies can capture in deploying ultra-low latency networks. Network operators or emerging players need to establish coherent business models and strategies based on partnerships with the industry verticals in order to create value in their service delivery.

Table \ref{tab:Market_findings} summarises the main findings of this market analysis and further details are given in the following subsections. For each vertical, three market components are presented: the market size, stakeholders in the business ecosystem, and opportunities for telecom players.

\begin{table*}
	\centering
	\caption{Market Size of Low Latency Applications in 5G}
	\label{tab:Market_findings}
	\begin{tabular}{l l l}
		\toprule
		\textbf{Industry Vertical} 	& \textbf{Market segment size} & \textbf{Remarks} 	\\
		\midrule
        \midrule
		 {Healthcare} & {\pbox{8cm}{GBP 43b 2018 global\\
GBP 2.9b 2018 UK\\
66\% digital health\\
18\% Tele-healthcare\\
35\% mHealth (mobile apps and wearables)\\
Medical robot systems global market of GBP 15b}} & {\pbox{8cm}{10\,ms Round trip delay to support haptic feedback\\
Tendencies to evolve towards remote applications\\
Mutual gain in merging tele-healthcare with connected home \\
50\% Potential Market share to network operators}} \\
		\midrule
		{Automotive} &
			{\pbox{8cm}{GBP 104.2 billion by 2019 connected car global:\\
(a) driver assistance: GBP 42b\\
(b) safety technologies:   GBP 28b\\
(c) safety and autonomous driving: 66\%}} &  {\pbox{8cm}{10\,ms One way delay in Cooperative driving\\
36m cars with SIM cards to be sold by 2018\\
Partnerships between Telecoms and Manufacturers\\
Global revenue of  GBP 3.4b for telecom companies}} \\
		\midrule
		{Entertainment} &
			{\pbox{8cm}{Media and Entertainment UK Market:  GBP 85.32b\\
Global AR and VR market is GBP 118.5b\\
Driver for VR is gaming with 76\% of content in gaming\\
Gaming in the UK:  GBP 3.9b consumer spend in 2015\\
Mobile Gaming in UK: GBP 548m}}& {\pbox{8cm}{7\,ms Round trip delay to support VR/AR\\
Largest industry in wearable segment growth\\
VR and AR are driven by Gaming industry\\
Revenues are dependent on level of added value provided}}\\			
		\midrule
		{Manufacturing} &
			{\pbox{8cm}{ICT Sector in Manufacturing market (EU): GBP 755.7b\\
Robotics and autonomous system technologies:\\ GBP 13b by 2025 in the UK \\
Ethernet and Wireless growing in industrial automation}} & {\pbox{8cm}{Sub-1\,ms One way delay in control applications\\
Ethernet solutions are growing fast \\
Cyber-physical systems to play primary role\\
Opportunities in low latency and data analytics}} \\
		\bottomrule
	\end{tabular}
\end{table*}
\subsection{Remote healthcare and medical intervention}

Mobile health (mHealth) can be defined as the use of mobile communication and devices for providing healthcare services such as remote monitoring, remote diagnosis, prevention and treatment. The applications can be divided into two general areas\cite{Deloitte}; one relates to the support and assistance provided at a distance using ICT, such as fall alarms or reminders of medicine intake. The second area is related to the remote exchange of clinical data between a patient and their clinician, such as continuous monitoring of chronic condition.
These two general areas represent great potential to improve the delivery of health service and evolve into more connected heath systems. On a longer-term perspective, the acceptance of ICT in healthcare delivery is expected to lead to more critical use cases such as remote interventions; however, there is still no market research study that addresses the topic of remote interventions where there is a strong need for ultra-low latency networks.

\textbf{Mobile health global market size:}
The global market for digital health was worth GBP 23 billion in 2014 and is expected to almost double to GBP 43 billion by 2018, according to\cite{Deloitte}. In 2014, the UK market size was GBP 2 billion and forecasts show a growth up to GBP 2.9 billion by 2018 with a Compound Annual Growth Rate (CAGR) of 11\%. This growth is mainly given to the increased interest of mobile health apps and health analytics. Overall, the UK digital health market represents approximately a 9\% share of the global market in 2014. Since the UK is recognised to be a developed market for mobile and digitised health systems, it can be used as an example to study how the market is segmented. The overall digital health industry market in the UK can be distributed as follows \cite{Deloitte}:
\begin{itemize}
	\item Digital health systems, which comprises electronic versions of traditional paper records, represent the largest market both globally and in the UK, where they contribute 66\% of digital health sales. This market is dominated by large, often international companies. Market penetration is already high in UK primary care as there have been several government initiatives that have driven the adoption of digitalization.
	\item Tele-healthcare, which includes remote monitoring, remote diagnosis, prevention and treatment, is the second largest sub-sector and contributes 18\% of the UK digital health market, and it is predicted to grow at 17\% every year until 2018.
	\item The most promising market for growth is mobile health apps, which is currently the smallest digital health market sub-sector but is predicted to grow at 35\% in the UK and 49\% globally from 2014 to 2018, this market is strongly focused on the wellbeing perspective, with apps dedicated to lifestyle activities. It is expected to merge into more health and care apps in the future.
\end{itemize}

\textbf{Business ecosystem for mobile healthcare:}
The firms involved in mobile healthcare value creation according to \cite{GSMA_mHealth2} comprise device/chip manufacturer and vendors, software/application developers, network operators, healthcare professionals, healthcare system (i.e., insurance or governments), and patients. But it is challenging to define value chains in a market where complex business interactions take place. For instance, relatives of patients are many times direct beneficiaries and users of mobile applications as well.

Regarding technical roles, device vendors primarily offer equipment that gather body vitals and transmit them to back-end servers over mobile networks. Content and application players offer information-based services either in a stand-alone fashion or through tie-ups with mobile operators and other aggregators. Some have also developed applications that improve the efficiency of healthcare providers.

Many mobile operators are active players in offering mobile health services and offer solutions beyond simple connectivity services. For instance, many operators offer content-based wellness information services to consumers. Some also offer sophisticated end-to-end solutions aimed at improving the efficiency of healthcare systems. Others facilitate mobile tele-medicine and health call-centres by partnering with healthcare providers. They also provide real-time connectivity for devices as well as managed services for monitoring vital body parameters of patients.

\textbf{Opportunities for telecom players:}
According to the study done in \cite{GSMA_PWC}, mobile operators are expected to be key beneficiaries of the expected growth in the mobile health market and manage nearly 50\% share of the overall market. However, this expectation needs to be matched with a willingness to deal in a highly regulated and fragmented market \cite{Labrique2013}, and strategies should consider the integration with existing health system functions to complement the objectives of health systems \cite{Labrique2013}. Based on \cite{GSMA_PWC} the market opportunities for telecom players is divided into these service categories:
\begin{itemize}
	\item Monitoring services for chronic diseases and aging population are expected to result in the increase in the adoption of home monitoring services.
	\item Treatment services, which will mostly be delivered through apps in large scale in Europe by 2017.
	\item In the remote intervention use case, current estimates from Europe and Japan indicate that the market for robotic and autonomous systems products and technology, for non-military sectors, will be in the order of GBP 70 billion by 2020-2025.
\end{itemize}

Network operators have numerous opportunities to support mobile health solutions within hospitals, physician practices and through home healthcare providers. Traditionally operators are connectivity providers, and the most straightforward business model should be to create monthly plans for the consumer that connects mobile health apps with hospitals, or connectivity provision for hospitals for the mobile health devices. One step further, should be to provide added value services to the healthcare industry by adopting the role of application and solution provider, where the operators leverage on their strong infrastructures and can provide converged solutions that go beyond connectivity such as security, data access, cloud access, etc. A similar model could come from different partnerships made with vendors or device manufacturers and developers to sell bundled and closed solutions of mobile health directly to enterprises.

The 5G Infrastructure Association takes a systems’ integration proposition for business models to serve the transformation in the way of delivering healthcare that is brought by ICT. The suggestion is to work on flexible strategies that allow each actor to focus on its core competences. For this, the key is to provide interfaces between business roles \cite{5G-IA2015}.

\textbf{Implications:}
Studies done so far have very well addressed the global mobile health market, and have shown strong predictions for business opportunities for network operators. The strong position of network operators is recognised, not only as a connectivity provider, but also as a service provider, allowing to extend the current business opportunities.

According to \cite{Deloitte, GSMA_PWC} tele-healthcare have already started to merge with the mobile health apps market.
This convergence between wireless communication technologies and healthcare devices has started to reshape the health sector \cite{OCDE2015}.
Disruptive changes in the mobile health apps business models could impact the current model of delivery for healthcare, and tele-healthcare is likely to converge with mobile Health and connected home solutions. The long-term increase of public expenditure in connected health systems increase the possibility of integrating robotic and autonomous systems to support remote healthcare with enhanced tele-medicine applications. This increases the overall market opportunities by introducing robotics and autonomous systems solutions. For example, solely the robotic surgery market is expected to be worth GBP 15 billion globally by 2021.

Moreover, network operators and communication providers are in a position to support the needs of the healthcare industry, by providing remote monitoring and access to medical data. The main challenges are related to the generation of an integrated value proposition and the roles that operators could take. These roles can span from just connectivity providers to system integrators \cite{Foh2012}.
The great value introduced by 5G that allows to increase market size and opportunity, is the support of remote monitoring scenarios, where ultra-low latency applications are more than likely to be considered in a near future. The market for remote interventions, such as remote surgery, is still in a nascent stage and it is not possible to foresee the implications.
\subsection{Assisted Driving and Transport Services}
Low latency networks are required in the automotive industry mainly to support enhanced assisted driving and autonomous driving.

\textbf{Connected car global market description and size:}
In the context of the self-driving car, by 2050 nearly all vehicles are expected to be self-driving cars. Study in \cite{GSMA_ITS} underlines that nearly 54\,million self-driving cars will be in use in the world by the year 2035. According to \cite{GSMA_ITS, GSMAAutomotive} the global connected car market has been forecasted to reach GBP\,104.2 billion by 2019, growing 34.7\% per year in the period of 2013 to 2019. Similarly, work in \cite{PWC_Connected} has recognised that the added revenue of the connected car market will increase up to GBP 95 billion by the year 2020, mainly driven by driver assistance and safety technologies, both with an overall value of GBP 42 billion and GBP 28 billion respectively by the year 2020.

Studies in \cite{PWC_Connected, PWC_Connected2} divide the overall connected car market into seven product categories, which will therefore distribute the market as follows:
\begin{itemize}
	\item Autonomous driving, includes all the vehicle operations that are done without human interaction. There is high potential growth of this segment, and it is recognised to be 33\% annual revenue growth being the fastest growing connected car segment.
	\item Safety, includes technologies that encompass external warnings for drivers, like weather, road conditions and emergency functions (such as the emergency call system). Safety and autonomous driving are the largest categories with 61\% of the total market share.
	\item Mobility management, are those functions that improve mobility and safety in an efficient way, by empowering fuel-efficient routes, for example. Technical requirements for these systems are currently available in the market and it is widely used in today's cars, therefore the growth rate will slow in the next years to a 5\% annual increase in the period 2015-2020.
	\item Vehicle management, refers to the monitoring services to reduce the car's operating costs and improve the overall information management. The increase in this segment is forecasted to 15\% annual increase until 2020.
	\item Entertainment, the revenue of this market is forecasted to increase 18\% annually.
	\item Well-being, includes services that address the comfort and safety of the driver in particular, such as anti-fatigue, music or climate, and will be extended in a near future to vital functions monitoring. This segment is expected to grow 31\% annually, mainly driven by the elderly drivers.
	\item Driver assistance, are those technologies that can improve the performance of the car. Technology is advancing quite fast in this segment, since many car manufacturing companies are investing huge efforts in the R\&D of new and sophisticated solutions, such as: platooning, assisted take over, see through. By the near future (i.e., 2017) these technologies are going to be widely available, and by the year 2020 these systems may not require any intervention or monitoring of the driver. This technological run drives a 40\% of annual growth in this segment, being the segment with highest growth in the connected car market. There is still uncertainty in the legal and regulations in this segment, especially regarding the fully autonomous driving.
\end{itemize}
The overall connected car market is driven by the consumer engagement with all these new products, possibly triggered by the user's need for connectivity. Also, according to \cite{GSMA_ITS}, the rapid advancements in the networks have been a clear opportunity to boost the in-car connectivity technological solutions, which has in great extend helped the growth of this market. This same study also suggests that the connected car penetration will increase globally to 60\%.

\textbf{Business ecosystem for connected cars:}
The main players in the connected car value chain are \cite{MCKC}:
\begin{itemize}
	\item Vehicle Manufacturers: these are mainly the automotive suppliers and the manufacturers. They are very focused in capturing new revenue opportunities, by extending their hardware services to offer integrated software services for infotainment, navigation and safety applications.
	\item Digital Players: these companies are adapting the user smartphone needs into the car by focusing on car-specific needs.
	\item Telecom Players: new opportunities have been identified in terms of infrastructure. Connected cars can provide huge business revenues for mobile operators, as more than 36\,million cars with pre-installed SIM cards are estimated to be sold by the year 2018, there is a potential global revenue opportunity of GBP 3.4 billion for telecom companies.
	\item Automotive Insurers: they are increasing their business opportunities by offering extended coverage for the telematic-based applications.
	
\end{itemize}
The emerging technologies such as 5G and the evolution of complementary industries (such as digital players) have increased the number of potential services that can be offered \cite{GSMA_CCBM}. There has been a change of paradigm in the definition of the business roles between the different players of the connected car industry, but nearly all market analysts agree that there is a strong need for strategic partnerships, in particular cooperation between the telecom players and the car manufacturers.

\textbf{Opportunities for telecom players:}
Network operators in general play an instrumental role in the development of the connected car services. In particular, global telecommunication alliances have well studied the network operator's opportunities in the connected car market, and it is agreed in the communities that the operators can leverage strongly in many of today's core activities and bring value to the value chain \cite{GSMA_CCBM, ConneCarArticle}. Telecom companies are critical enablers in aspects such as billing, which is one strong capability operators bring, and it is currently one of the critical aspects of the connected car industry. Other aspects where telecom operators can improve are device or subscription management, controlling upgrades or roaming information, or providing information related to the location to enhance services. Other big aspects are the Telematics Service Provider (TSP) platforms, which are expected to gain around 11\% of the connected car market share, and many telecom companies are expanding their presence, operators can take care of maintenance and service upgrades. Operators can handle big amounts of data that can be analysed to provide added value in services, such as traffic patterns based on mobile users’ position, vehicle parameters, or driving coaching services for inexperienced drivers. This makes them strong players to provide cloud solutions or data collection and analytics services, market opportunity that requires collaboration agreements with the automotive or insurance industries. Finally, as general connectivity providers for consumers there are numerous models that can be adopted, as for example offering content and storage services for in-car entertainment or providing wireless connectivity inside the car, and facilitate the machine to machine (M2M) communication to collect data from sensors or wearables.

\textbf{Implications:}
Network operators have opportunities enabling applications inside the connected car ecosystem that do not necessarily involve the use of low latency networks: cloud-based solutions, car-occupancy measurements for toll lanes, authentication and security and car sharing are some examples widely described in the literature \cite{GSMAAutomotive, GSMA_CCBM}. The cooperation with car manufacturers enables to complement each other’s needs since mobile technology is in great extent one of the main drivers of the change on the way people use cars nowadays and in the future. Based on this, mobile operators can play an important role in four key elements of smart city services, which are also directly applicable in the connected car context:
\begin{itemize}
	\item Connectivity: network coverage and management, multiple technology aggregation, etc.
	\item Data aggregation/analysis: combining and optimizing data delivery, reducing traffic overload in the network
	\item Service delivery: delivering real-time information to people and machines that will enable them to adapt and respond to events in the city;
	\item Security: operators are experienced in handling security with sensitive data.
\end{itemize}

In general, operators and car manufacturers have complementary roles, which increase even further the opportunities to succeed if strong relationships are built between both players. Currently, mobile operators can bring expertise in providing connectivity and support in the connected car service provisioning, areas where manufacturers are not well positioned. In the long run, the need for connectivity support with the use of ultra-fast networks is going to be a requirement for operators offering connected car services, since future service models are evolving towards autonomous driving, enhanced safety and security information, and improved driver assistance. This will require strong critical M2M support and good device interoperability.
\subsection{Entertainment: Content Delivery and Gaming}
Online entertainment represents a wide market that covers from TV and audio-visual industries in general to video games industries. As seen in previous sections, there is an interest across in the entertainment industries to provide the end user with more immersive experiences, which will require low latency networks. Wearables are key into bringing a full immersive experience, this could be either in the form of VR or AR goggles, 3D sound headsets or haptic devices.

\textbf{Entertainment market description and size:}
Based on the forecasts shown in \cite{ITA}, the overall media and entertainment UK market will expand to GBP 85.32 billion in 2018, with two main growing sectors: digital video entertainment and video games, which are both growing exponentially. Ofcom research in \cite{OfcomComs} shows that the online TV market revenue in the UK has increased from GBP 102 million in 2009 to GBP 908 million in 2014, mainly driven by the consumers’ preference to use online services to watch TV and films, and an increased adoption of online TV and video services.

Moreover, consumers are expecting from wearables to improve the already existing offer in entertainment. Work in \cite{PWC_Wearables} states that 73\% of people expect that entertainment is going to be more immersive and fun with the use of wearables.
In particular, the global augmented and virtual reality market is forecasted to be worth GBP 118.5 billion by 2020, being the smallest share virtual reality, with GBP 23.7 billion. Virtual reality will grow mainly in games and 3D films, due to its full immersion it is not possible to consider it to be used everywhere. On the other hand, AR has increased potential in markets beyond games, from enterprise applications, such as real estate, to enhanced outdoor experiences. Currently, the key driver of VR is the gaming industry, since games make up 76\% of all VR content, 37\% for PC, 32\% for console and 7\% for mobile, according to the UK interactive entertainment fact sheet in \cite{UKIE}.

In the context of gaming, the UK is estimated to be the sixth largest market in the world, and it is estimated to be worth GBP 3.9 billion in consumer spend in the year 2015 \cite{UKIE}. The UK digital market is divided in: GBP 741 million for PC, GBP 118 million for console and GBP 548 million for mobile. According to \cite{Comviva} online PC games and mobile games are the fastest growing segments in the digital gaming industry.

\textbf{Business ecosystem in entertainment:}
Traditional broadband and media value chain has been commonly classified as:
\begin{itemize}
	\item Content creation: original digital content that can be used to augment the reality, virtual reality broadcast, or any immersive content.
	\item Content aggregation and publishing: providing the publishing environments or platforms.
	\item Content distribution: delivering content to the end user.
	\item Content consumption: customer experience.
\end{itemize}

However, the overall entertainment business ecosystem is changing rapidly. In content delivery, there is no linear value creation anymore and digital content is not bounded by the constraints of traditional media, and content creators can decide to take it directly to the consumer via digital platforms, as for example YouTube or StreamCloud. Moreover, content aggregators and distributors are now generating content themselves, as is the case with Netflix, or buying rights as happens with sports. Also new players need to be considered which are the key enablers of the fully immersive experience, new hardware and software players, which in fact can bring some similarities to the wearables value chain.

Network operators have a key role as enablers of the fully immersive experience \cite{VRArticle}, and the main reason is that to be able to support these innovations in online platforms connectivity providers need to ensure they can cope with the capacity and latency demands. Intelligent traffic management solutions, compression algorithms, investments in ultra-low latency and high-throughput networks will help operators to cope with the demands of VR content.

\textbf{Opportunities for telecom players:}
Entertainment companies have the largest opportunity for growth in the wearable technology market, and currently there is a high number of applications available \cite{PWC_Wearables}. Overall, new business models for network operators need to be focused into adding value to the services provided, which means to collaborate closely with suppliers to ensure good end-to-end service quality that preserves a certain level of customer experience. In this line, 5G partnership alliances underline that collaborative services imply the association of fixed and wireless as well as terrestrial and satellite network service providers, to deliver services with a common multi-service control layer and assured quality \cite{5GPPWPMedia}. In this sense, converged operators have a competitive advantage.

On the other hand, following the analysis on business models for mobile gaming done in \cite{Comviva}, the online gaming model, which allows users to play over the operator's network (no need to download), has the potential to increase revenues and opportunities in this market. The particular business model, can be extended to any entertainment service, and it considers:
\begin{itemize}
	\item Operator builds a vibrant entertainment brand and reinforces customer loyalty by offering improved user experience.
	\item Reduce dependency on the device by exploiting benefits of cloud based gaming irrespective of the device (PC or mobile device).
	\item Ubiquity at home or on the move, converged plans that tackle consumers that use entertainment at home or on the move indistinctly.
	\item Game advertising, allowing for richer and more innovative ways of providing advertising.
\end{itemize}

\textbf{Implications:}
The entertainment industry has derived into multiple markets with certainly multiple opportunities for network operators. In this sense, providing fully immersive entertainment experiences to the customer enables opportunities for the operators to deliver high quality services. The overall wearables market is growing fast, and in the entertainment industry those having major impact are VR and AR, which also pose strong latency and capacity requirements in the mobile network. However, new business models arise for mobile and fixed network operators to provide added value into the services. As well, we have shown that there is an opportunity to grow in the B2B model, since operators can not only play a primary role in the service/content delivery, but also in the support of the entertainment industry providing cloud based secure platforms, and build strong partnerships with suppliers.
\subsection{Industry Automation}
Europe has been an early adopter of the digital innovation in the manufacturing sector. Nevertheless, there are incentives to transform the manufacturing processes in the future with the use of intelligent networks that allow to remotely control or interact in the product development and production.

\textbf{Industrial automation market description and size:}
According to the European commission (EC) \cite{EUWhitePaper}, the overall market of the EU manufacturing corresponds to the 15\% of its GDP, about GBP 1890 billion, and the ICT sector accounts for GBP 755.7 billion of it. Europe has 30\% of the world market share in robotics and factory automation and 33\% in embedded digital systems, enterprise and product design software.

However, there has been a 1.3\% drop of the GDP share since 2008. In response to this decline, the EC has set a target that manufacturing should represent 20\% of the GDP by the year 2020. This target comes together with investment requirements, and to realize the factory of the future concept additional GBP 84.1 billion of investments are estimated to be needed. Industries with state of the art production processes are far more competitive, can afford higher margins and pay off their capital needs. This means that capital investments are going to result in higher profitability. Therefore, with regard to the market perspectives, market experts believe that in five years more than 80\% of the companies will have digitized their value chain, thereby increasing 18\% in production efficiency. Overall, digitized products and services generate \EUR{110} billion of additional revenues in the European industry.

To support many of this future enhancements in the industrial manufacturing business, remote applications with the use of robotics need to be empowered by the industry, it is estimated that optimising current robotics and autonomous systems technology would raise productivity in manufacturing by up to 22\%.

\textbf{Business ecosystem in industry automation:}
The main idea of the FoF or industry 4.0 is to change the way the industrial value chain is conceived. The new concept will represent a new level of organization and control of the production chain. While traditional players such as logistical partners, suppliers of parts, suppliers of sensors and actuators, and manufacturers themselves, are going to still be playing a major role in the industry business, new players will enter the scene as suppliers of cyber-security, suppliers of data storage and management, suppliers of connectivity, suppliers of specialized technology, and suppliers of automation systems. Nowadays, the majority of players expect new competitors to enter the market, 84\% of technology suppliers expect new competitors and 58\% of manufacturers expect new competitors entering the market \cite{Industry4}.

Telecommunications companies are going to have an instrumental role in this new industrial revolution. Study in \cite{Industry4} recognises that many of the disruptive technologies are driven by small, innovative companies that have specialized in a given field, and it is likely to have an increasing emergence of highly specialized players specially in the telecoms area, providing solutions for data, connectivity and security.

\textbf{Opportunities for telecom players:}
The disruptive change in the value chain is going to enable new business models for the telecom players. One of the main drivers of the industry change is data and leaders across industries are leveraging data and analytics to achieve a step change in value creation. According to \cite{Industry4} a big data and advanced analytics approach can result in a 20-25\% increase in production volume. Thus, new business models arising are centred in adding value around the whole connected items and collect, use and share data. Specifically, offering solutions around integration and new services, enabling manufacturing companies to capture this emerging value.

Business models particularly interesting for network operators are around providing technology platforms and data-driven services. The role of 5G connectivity is also going to be instrumental in enabling most of the applications described in previous sections, and apart from including business models that add value to the connectivity services, network operators need to consider:
\begin{itemize}
	\item Ubiquitous communications for both machines and humans.
	\item Wireless and fixed system integration in a seamless way.
	\item Interoperability and unified network solutions.
\end{itemize}

\textbf{Implications:}
The major changes that are expected for the manufacturing industries strongly rely in the use of communications networks, which directly reflects the opportunity for network operators or other communication providers to engage in partnerships with other industries in the value chain to provide efficient and reliable end-to-end data and communications infrastructures. Telecom companies can leverage in many of their currents strengths to devise high quality business models that can add real value to the manufacturing business. Based on the market analysis and the estimated future for the industry 4.0 these opportunities are already engaging new competitors to appear with new business models, especially in the telecommunications industry.

\section{Discussion: the business transformation, new roles and market position for telecom players}
Ultra-low latency applications have intrinsic technical challenges, and they represent a major change in the telecom business domain. Similar changes have already been studied in the context of IoT and 5G systems; in this section, we present the main business research findings in the IoT industry and how these can be applied to ultra-low latency applications.

The role and market position of network operators will change as new ways to cooperate emerge in the future. For 3G and 4G networks, the cooperation between competing operators started to emerge in some countries. In these cases the mobile networks are jointly deployed and operated, the network resources are shared but the operators compete for the end-users \cite{Markendahl389689,Markendahl666147}. This so called \textit{network sharing} is a form of horizontal cooperation and competition among network operators.

As remarked in the previous section, for 5G systems we will see many more types of actors as well as new types of cooperation, which are essential to deliver low latency services. Hence, the discussion will no longer be about cooperation between two operators. Instead, we will see cooperation in larger business networks or value configurations \cite{CM2017,pujol2016mobile}. We will see an evolution of the operator-centric business model and cooperation patterns \cite{pujol2016mobile,laya2015tele}.
To provide ultra-low latency, the same as happened in IoT applications, mobile operators need to cooperate with actors providing sensors and other types of hardware, service platforms, applications and integration services. Single-firm business models need to be replaced by ecosystem-oriented business models taking into account the resources and capacity from the different actors in the business network \cite{Leminen2015}.

Many business researchers have looked into the multitude of actors in emerging industrial networks. The complexity of the ecosystem will be larger since both the number of involved actors as well as the type of actors increase \cite{Basole2009}. Work on networked business models and industrial networks claim that we need to go from firm-centric business models to network-embedded business models \cite{Palo2013,Bankvall2016}. Here, vertical cooperation and competition is a key concept \cite{Lacoste2012,CM2017}.

The simultaneous competition and collaboration has been presented in different ecosystem studies for different industries like energy \cite{7073816}, smart cities \cite{7552678,CM2017} and health \cite{markendahl2013business,VanMeeuwen2015}. The complexity increases with the number of actors. This is especially evident in the public sector, typically for health and public transport services, where societal actors naturally are part of the business networks. In the public sector, the multi-actor business models also have to take into account aspects like public funding, legislation and policies.

In the traditional telecom business, the roles and market position of network operators and Vendors of telecom equipment (Vendors) are very clear. The operators have the business relation with the end users through pre-paid or post-paid subscriptions, today typically including voice and messaging services and a monthly bucket of mobile data. The vendors support this operator business in two ways: \textit{(i)} by selling network equipment and \textit{(ii)} by providing \textit{managed services} meaning that the vendor take care of the network operation and maintenance (often also network deployment).
The core business here is the provisioning of connectivity services to consumers and businesses.

\begin{figure*}[t]
  \centering
  \includegraphics[scale=0.6]{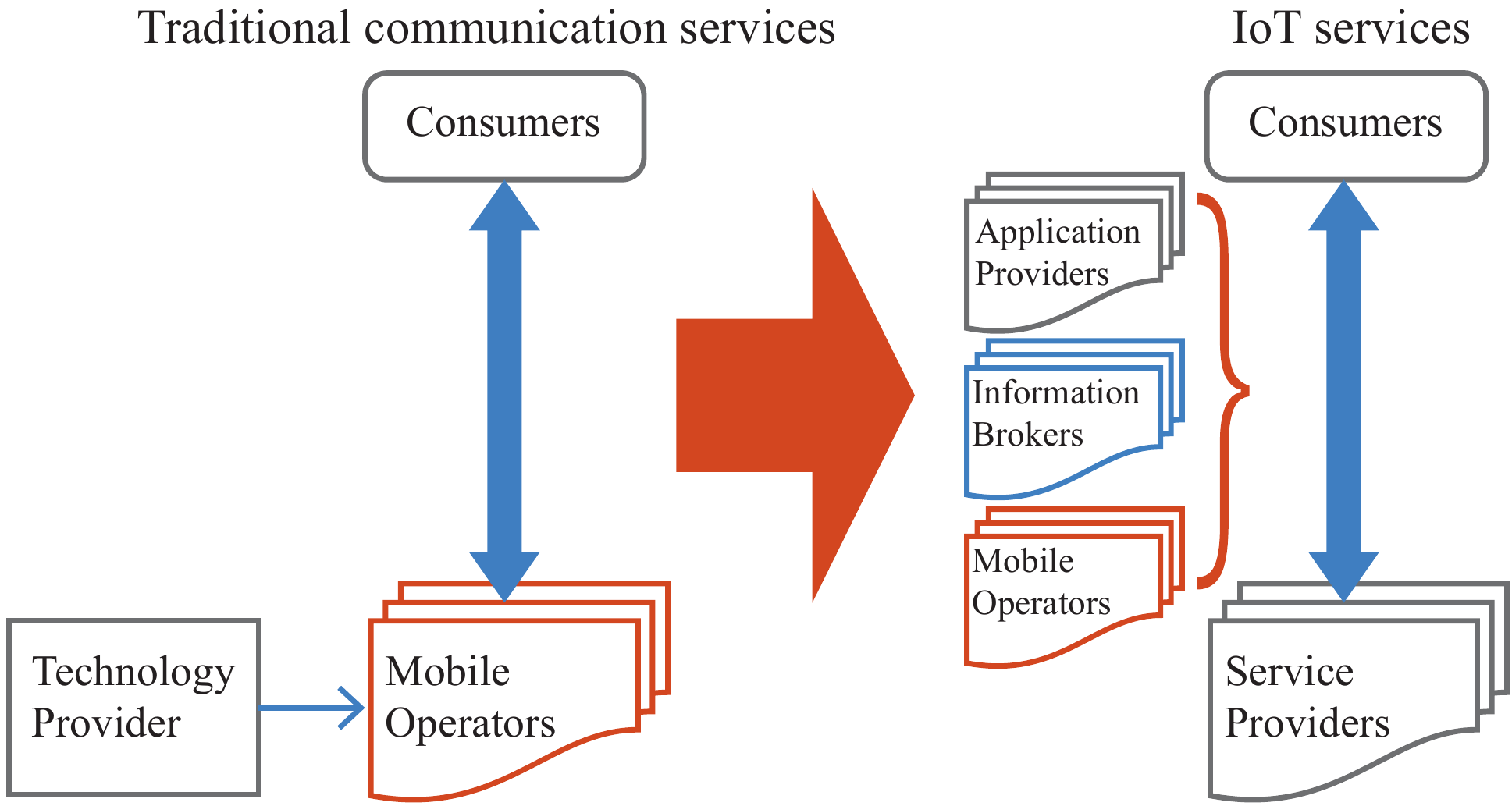}\\
  \caption{Change of roles and market position for traditional telecom actors going from connectivity to IoT services (figure adapted from \cite{markendahl2013business}).}\label{fig:businesschange}
\end{figure*}

Considering the future ultra-low latency applications, we can draw parallel lines with early IoT service developments. In the IoT context, when new services are introduced, the positions of telecom actors change completely compared to the traditional communication services. The IoT services are part of some core service within some of the industrial verticals. Hence, the core service is provided by some other actor that has the key business relation with the users. In this case, the role of the telecom actors is to support this core service offered by actors in the industrial verticals. This change of market position is illustrated in figure \ref{fig:businesschange}. In this re-positioning we can identify both cases where operators and vendors compete and when they collaborate.

\begin{figure}[t]
  \centering
  \includegraphics[scale=0.55]{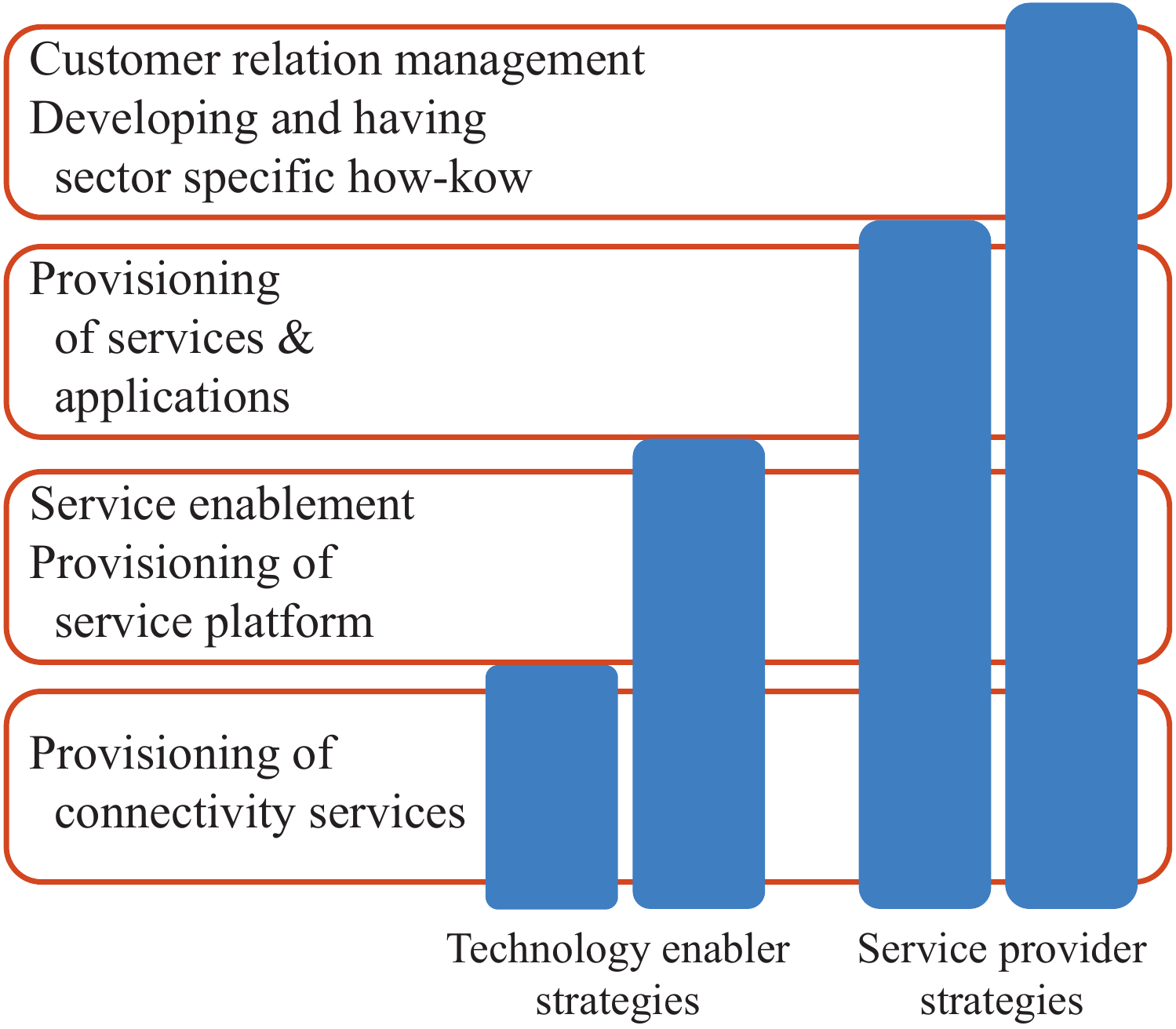}\\
  \caption{Different scope and strategies for Network Operators entering IoT business.}\label{fig:scopes}
\end{figure}

Continuing with lessons learned from IoT services, when network operators offer IoT services different strategies can be identified depending on how many roles the network operator will take as part of their service offering. This is related to what activities and resources the operators are able to control and have managerial responsibilities, as shown in figure \ref{fig:scopes}. One existing strategy is to only focus on technology enablers, typically referring to connectivity and service platforms. In this type of strategy the operator often builds a partner network in order to be able to offer a full offer. In the context of IoT these partners provide e.g. sensors, integration, application and sector specific know-how. Some of the Network Operator's partner can be responsible for the contact and contract with the provider of the core service.

Another Network Operator strategy is to offer a full service bundle under its own name and brand. It may be that some components, such as sensors, platform, integration or applications, are provided by a trusted partner, but then it is shown in the market under the operator's brand.

All in all, taking part in provisioning of ultra-low latency services (or IoT services) implies fundamental changes for both operators and telecom vendors. The market positions will change compared to those for connectivity services and the actors need to take new roles and develop new type of skills and business relationships. We argue that this trend will continue and be more evident in the emergence of low-latency applications. The same as in IoT services, customers are not always end users (as for connectivity services), instead, the customers for telecom actors can be service providers in different industrial sectors. These are often big companies or organizations which will have an impact on the operators' marketing, sales activities and organization. In addition, technical solutions are not the same as for traditional B2B connectivity services in the past, since the communication requirements for low-latency applications are diverse and application-specific.

\section{Conclusions}
A large number of industries are taking part of the requirement design of the future 5G network by introducing new applications. These industry verticals will pose new and strong challenges to network providers; in this paper we have thoroughly surveyed the requirements for ultra-low latency applications of four major industries and their specific applications: Healthcare, Automotive and Transport systems, Entertainment and Manufacturing. Among all the trends being considered in the industry verticals ultra-low latency is key when aiming for fully immersive applications, like robotic tele-operation, or the use of VR, and it is of course essential in critical and safety applications. One of the key arguments of this study is that current communications standards, such as 3GPP LTE Advanced or widely deployed fixed networks cannot meet these stringent requirements, which opens the door to a whole new set of innovative technologies that enable such low latency figures. Thus, the main reason 5G needs to support for ultra-low latency services is the roadmap to include such applications in a mobile network.
5G needs to be able to provide these services within certain levels of QoS, and some of the research is being driven towards the improvement of radio access network protocols or the introduction of network virtualization, which allows to efficiently place network functions and reduce the overall delay. Also, 5G needs to be a fully converged network, where multiple fixed and mobile access technologies can be flexibly selected sharing core network functionalities, leading to latency and reliability improvements. To highlight this, we have included a review of the main trends in technology to tackle the 5G latency requirement.

However, all these efforts in creating and deploying a low latency network need to be equally reflected in a strong market need. In this line, we have surveyed the main market research findings for each industry vertical. As an outcome of this study, it is clear that the operator will take an instrumental role in delivering these services or applications, and the specific low latency revenue largely depends on the ecosystem-oriented business models selected by the operator and the specific market addressed. Support for low latency applications in the entertainment industry show a clearer opportunity map, since the support for online VR and AR related services are strictly related to the availability of a network capable of delivering such experiences. The other vertical in similar situation is the healthcare industry, which needs for connectivity solutions to extend remote applications on the move and at home. The automotive and transport vertical has increased engagement from the telecom industry in terms of standardisation efforts, communication trials and strong partnerships. Industrial manufacturing is a less mature vertical, since it has telecom players with very specific business models or proprietary solutions that have been developed to satisfy its specific needs. In this sense, business model research is required to find out just how to best enter the market (and compete with the providers of proprietary solutions) from a network vendor and operator perspective.

Answering our initial research question: \textit{will it be cost effective for telecom players to build ultra-low latency in 5G networks?}, we can say that telecom players can clearly benefit by supporting 5G low latency applications, provided new business models are created, and the community works together building strong partnerships and allow to co-create technology that can match the specific needs of each industry vertical. The choice of new business models will determine the economic effect of 5G and the revenues of low-latency services.

\section*{Acknowledgments}
The authors would like to thank Dr. Peter Marshall for his valuable contributions to this work. This work is supported by the Ericsson 5G and Tactile Internet industry grant to King's College London, the BT collaboration with King's College London and Ericsson, as well as the Sweden's Innovation Agency (VINNOVA) project on IoT Ecosystems.

\bibliographystyle{IEEEtran}
\bibliography{Bibliography}

\end{document}